\documentclass{article}

\usepackage{arxiv}

\usepackage[utf8]{inputenc}
\usepackage[numbers,comma,square]{natbib}
\usepackage[figuresleft]{rotating}
\usepackage{txfonts}%
\usepackage{multirow}
\usepackage{array}
\usepackage{xcolor}
\usepackage[hidelinks]{hyperref}

\newcolumntype{L}[1]{>{\raggedright\let\newline\\\arraybackslash\hspace{0pt}}m{#1}}
\newcolumntype{C}[1]{>{\centering\let\newline\\\arraybackslash\hspace{0pt}}m{#1}}
\newcolumntype{R}[1]{>{\raggedleft\let\newline\\\arraybackslash\hspace{0pt}}m{#1}}

\definecolor{myML}{HTML}{08519C}
\definecolor{myOR}{HTML}{3182BD}
\definecolor{myNET}{HTML}{6BAED6}
\definecolor{myPS}{HTML}{BDD7E7}
\definecolor{mySTAT}{HTML}{EFF3FF}

\newcommand{\ML}{\textcolor{myML}{\blacksquare}}
\newcommand{\OR}{\textcolor{myOR}{\blacksquare}}
\newcommand{\NET}{\textcolor{myNET}{\blacksquare}}
\newcommand{\PS}{\textcolor{myPS}{\blacksquare}}
\newcommand{\STAT}{\textcolor{mySTAT}{\blacksquare}}

\title{A Review on Flight Delay Prediction}

\author{
 Alice~Sternberg \\
 CEFET/RJ\\
 \And
 Jorge~Soares \\
 CEFET/RJ\\
 \texttt{jorge.soares@cefet-rj.br} \\
  \AND
Diego Carvalho\\
 CEFET/RJ\\
 \texttt{dcarvalho@ieee.org} \\
  \And
Eduardo~Ogasawara\\
 CEFET/RJ\\
 \texttt{eogasawara@ieee.org} \\
}

\begin{document}
\maketitle

\begin{abstract}
	Flight delays hurt airlines, airports, and passengers. Their prediction is crucial during the decision-making process for all players of commercial aviation. Moreover, the development of accurate prediction models for flight delays became cumbersome due to the complexity of air transportation system, the number of methods for prediction, and the deluge of flight data. In this context, this paper presents a thorough literature review of approaches used to build flight delay prediction models from the Data Science perspective. We propose a taxonomy and summarize the initiatives used to address the flight delay prediction problem, according to scope, data, and computational methods, giving particular attention to an increased usage of machine learning methods. Besides, we also present a timeline of significant works that depicts relationships between flight delay prediction problems and research trends to address them.

The published version of this paper is made available at \url{https://doi.org/10.1080/01441647.2020.1861123}.

Please cite as: 

L. Carvalho, A. Sternberg, L. Maia Gonçalves, A. Beatriz Cruz, J.A. Soares, D. Brandão, D. Carvalho, e E. Ogasawara, 2020, On the relevance of data science for flight delay research: a systematic review, Transport Reviews

\end{abstract}

\section{Introduction}
\label{Introd}

Delay is one of the most remembered performance indicators of any transportation system. Notably, commercial aviation players understand delay as the period by which a flight is late or postponed. Thus, a delay may be represented by the difference between scheduled and real times of departure or arrival of a plane \cite{wieland_limits_1997}. Country regulator authorities have a multitude of indicators related to tolerance thresholds for flight delays. Indeed, flight delay is an essential subject in the context of air transportation systems. In 2013, 36\% of flights delayed by more than five minutes in Europe, 31.1\% of flights delayed by more than 15 minutes in the United States, and 16.3\% of flights were canceled or suffered delays greater than 30 minutes in Brazil \cite{Eurocontrol_coda_2017,anac_agencia_2017}. This indicates how relevant this indicator is and how it affects no matter the scale of airline meshes.

Flight delays have negative impacts, mainly economic, for passengers, airlines, and airports. Given the uncertainty of their occurrence, passengers usually plan to travel many hours earlier for their appointments, increasing their trip costs, to ensure their arrival on time \cite{balakrishna_accuracy_2010,fleurquin_trees_2014}. On the other hand, airlines suffer penalties, fines and additional operation costs, such as crew and aircrafts retentions in airports \cite{britto_impact_2012,tu_estimating_2008,evans_quantifying_2004,hsiao_air_2005}. Furthermore, from the sustainability point of view, delays may also cause environmental damage by increasing fuel consumption and gas emissions \cite{pejovic_tentative_2009,ryerson_time_2014,rebollo_characterization_2014,krstic_simic_airport_2015,balaban_dynamic_2017, xu_maximizing_2017}.

Delays also jeopardize airlines marketing strategies, since carriers rely on customers' loyalty to support their frequent-flyer programs and the consumer's choice is also affected by reliable performance. There is a identified relationship between levels of delays and fares, aircraft sizes, flight frequency and complaints about airline service \cite{daniel_when_2008,manley_impact_2008,bhadra_you_2009,pai_factors_2010,zou_flight_2014}. The estimation of flight delays can improve the tactical and operational decisions of airports and airlines managers and warn passengers so that they can rearrange their plans \cite{dariano_aircraft_2012}.

To better understand the entire flight ecosystems, vast volumes of data from commercial aviation are collected every moment and stored in databases. Submerged in this massive amount of data produced by sensors and IoT \cite{mellat-parast_linking_2015,campanelli_modeling_2014,
	mueller_analysis_2002}, analysts and data scientists are intensifying their computational and data management skills to extract useful information from each datum. In this context, the procedure of comprehending the domain, managing data and applying a model is known as Data Science, a trend in solving challenging problems related to Big Data.

Under this data deluge scenario, this paper contributes by presenting an analysis of the available literature on flight delay prediction from Data Science perspective. It seeks to summarize the most researched trends in this field, describing how this problem is addressed and comparing methods that have been used to build prediction models. This becomes more relevant as we observe an increasing presence of machine learning methods to model flight delays predictions. This analysis is conducted by establishing a flight delay research taxonomy, which organizes approaches according to the type of problem, scope, data issues, and computational methods. The paper also contributes by presenting a timeline of major works grouped by the kind of flight delay prediction problem.

Besides this introduction, the rest of this paper is structured as follows. Section \ref{FlightDS} introduces the flight delay scenario, describing a typical operation of a commercial flight, kinds of delays and their impacts. It also structures three different ways for treating the prediction problem. In Section \ref{Taxonomy}, a taxonomic analysis of the prediction is presented, showing the most researched topics, the scope of application, data and methods that authors are using to predict flight delays. Section \ref{Results} discusses the main results based on a timeline of publications grouped by the types of problems and their intersections. Finally, Section \ref{Conclusion} concludes our analysis by presenting major highlights and trends about delay prediction problem.

\section{The flight delay scenario}
\label{FlightDS}

\begin{figure}[!ht]
	\centering\includegraphics[width=\linewidth]{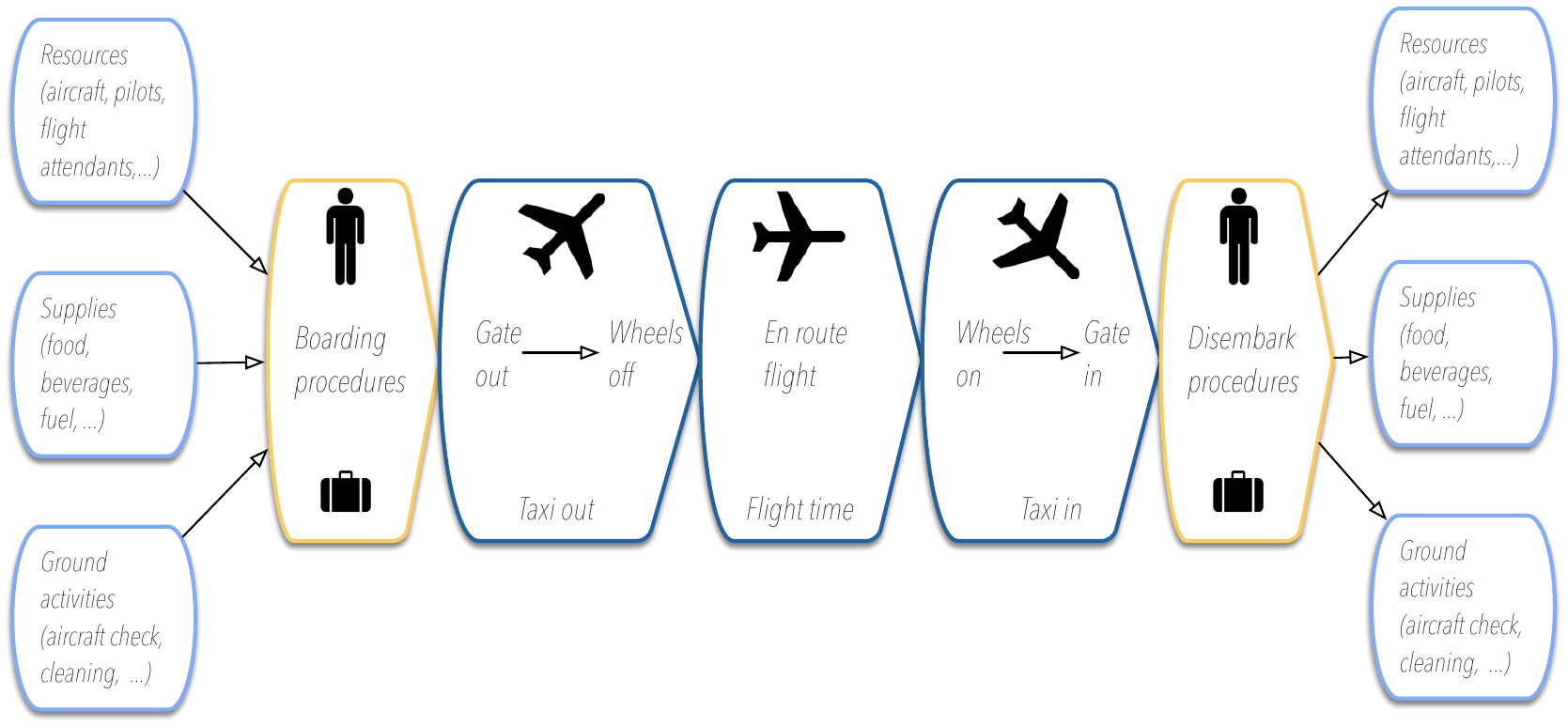}
	\caption{A typical operation of a commercial flight}
	\label{fig:typical}
\end{figure}

Commercial aviation is a complex distributed transportation system. It deals with valuable resources, demand fluctuations, and a sophisticated origin-destination matrix that need orchestration to provide smooth and safety operations. Furthermore, individual passenger follows her itineraries while airlines plan various schedules for aircrafts, pilots and flight attendants. Figure \ref{fig:typical} illustrates a typical operation of a commercial flight. Stages can take place at terminal boundaries, airports, runways, and airspace, being susceptible to different kinds of delays. Some examples include mechanical problems, weather conditions, ground delays, air traffic control, runway queues and capacity constraints \cite{reynolds-feighan_assessment_1999,hunter_advanced_2007,ahmadbeygi_analysis_2008}.

This scheme is repeated several times throughout the day for each flight in the system. Pilots, flight attendants and aircrafts may have different schedules due to legal rests, duties, and maintenance plans for airplanes. So, any disruption in the system can impact the subsequent flights of the same airline \cite{abdelghany_model_2004}. Moreover, disturbances may cause congestion at airspace or other airports, creating queues and delaying some flights from other carriers \cite{schaefer_flight_2001,xu_estimation_2005}. In this way, the prediction of flight delays is an essential subject for airlines, airports, Air Navigation Service Providers (ANSP), and network managers, like FAA \cite{faa_federal_2017} and Eurocontrol \cite{eurocontrol_european_2017}.

The flight delay prediction problem can be treated by different points of view: (i) delay propagation, (ii) root delay and cancellation. In delay propagation, one study how delay propagates through the network of the transportation system. On the other hand, considering that new problems may happen eventually, it is also important to predict further delays and understand their causes. Such occurrences, in this paper, are named as a \emph{root} delay problem. Finally, under specific situations, delays can lead to cancellations, forcing airlines and passengers to reschedule their itineraries. So, researchers focused on cancellation analysis try to figure out which conditions lead to cancellations. Moreover, it explores the airlines' decision-making process for choosing the flights to be canceled.

\section{Taxonomy}
\label{Taxonomy}

\begin{figure}[h!t]
	\centering\includegraphics[width=0.9\linewidth]{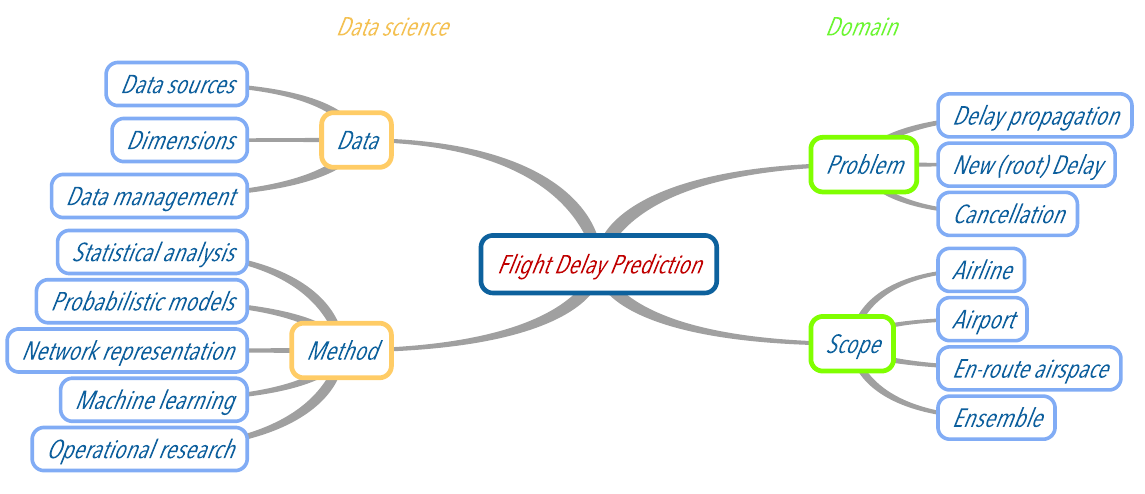}
	\caption{Taxonomy of the flight delay prediction problem}
	\label{fig:taxonomy}
\end{figure}

The main problems related to flight delay prediction are identified and organized in a taxonomy. It includes scopes, models, and ways of handling flight delay prediction problem. It considers flight domain features, such as \emph{problem} and \emph{scope}, and Data Science perspectives, such as \emph{data} and \emph{methods}. Figure \ref{fig:taxonomy} depicts the entire taxonomy while next subsections describe each component of the taxonomy and related work.

Regarding the available literature on flight delay prediction, we have conducted a systematic mapping study. The search expression string \emph{ (``airport delay" $\vee$ ``flight delay") $\wedge$ (``predict" $\vee$ ``forecast" $\vee$ ``propagate")} was used to query Scopus on October 2017. Query result brought 310 references. Additionally, 29 works were added using snowballing search.

We have selected 134 to build this review due to their relevance and direct link with the flight delay prediction problem. The main criteria to be included is to have the word ``delay" in the abstract, and the paper should have at least the one citation at Google Scholar per year before 2016. It means that to include a paper of 2015, it must have at least one citation, and so one. 

From this study, we were able to present a taxonomy that drives the organization of the following sections.

\subsection{Problem}
\label{Problem}

Problem is the core feature in domain taxonomy. As seen in Section \ref{FlightDS}, there are three major concerns regarding the flight delay prediction problem: delay propagation, root delay and cancellation. Depending on the emphasis of the research, authors select one of these lines to develop their models.

\subsubsection{Root delay and cancellation}
\label{DelayI}

Considering that new delay (root delay) may happen eventually, these root delays impair the performance of transportation network. Researchers create prediction models to tackle root delay, predicting when and where a delay will occur and what are its reasons and sources. This includes models that efficiently seek to estimate the number of minutes, probability or level of delay for a specific flight, airline or airport. 

A relevant number of works focused on predicting and estimating delay duration \cite{rebollo_characterization_2014}. Some approaches considered probabilistic models and innovation distribution \cite{mueller_analysis_2002, tu_estimating_2008}, whereas others find conditions for the occurrence of a root delay, such as passenger demand, fares, flight frequency, aircraft size, and taxi-out time \cite{balakrishna_accuracy_2010, zou_flight_2012}.

Particular circumstances, such as weather conditions, acts of God, aircraft problems, may lead airlines to cancel flights. Besides, airlines may directly cancel a flight, when factors like seat occupancy or cost savings are taking into consideration \cite{le_optimum_2008,xiong_modelling_2013}. 

\subsubsection{Delay propagation}
\label{DelayP}

In delay propagation, the primary objective is to understand how delay propagates through airlines and airports based on the assumption that an initial delay has already occurred in the transportation system. A particular scenario happens when delays are spread to other flights of the same airline as chain reactions \cite{boswell_analysis_1997,beatty_preliminary_1998, abdelghany_model_2004, wong_survival_2012}. Under this situations, it is important to measure how stable and reliable carriers can be to recover from delay propagation \cite{wu_inherent_2005, duck_increasing_2012}. Also, a delay may continue to propagate due to the scheduling of critical resources or retentions in other airports \cite{hansen_micro-level_2002}.

When scheduled time for take-off or landing is not fulfilled, flights need new slots that may be unavailable. In this scenario, it is important to understand the effects that a root delay in flight may produce to both departure and arrival airports \cite{xu_estimation_2005, pyrgiotis_modelling_2013, hao_new_2014}. Such phenomenon may increase the number of flights at some period, generating capacity problems and queues.

\subsection{Scope}
\label{ScopeA}

Delays can be induced by different sources and affect airports, airlines, \emph{en route} airspace or an ensemble of them. For analysis purposes, one may assume a simplified system where only one of these actors or any combination of them is considered. It should be noted that any scope of application can be combined with any problem mentioned in Section \ref{Problem}.

Some work focused on airports to predict delays for all departs considered all airlines and \emph{en route} airspace indifferently \cite{schaefer_flight_2001, rebollo_characterization_2014}. Airports are also the focus when the objective is to investigate their efficiency based on delays of all carriers \cite{pathomsiri_impact_2008,kim_deconstructing_2013, pyrgiotis_modelling_2013,kim_impacts_2016}. On the other hand, only airlines are considered when comparing the performance of two airlines under the same conditions \cite{ahmadbeygi_analysis_2008}.

An ensemble of airport and \emph{en route} airspace were studied to understand the relationship between congestion and delays \cite{hunter_advanced_2007, morrison_effect_2008}. Others considered airports and airlines as well to evaluate capacity problems and airlines decisions \cite{tu_estimating_2008}. There are many possibilities to ensemble scopes. This becomes important when studying the dynamics of air transportation systems, mainly when targeting root delay. 

\subsection{Data}
\label{Data}

Three fundamental questions about data are: Where to find flight data? Which attributes should be considered? Is it possible to handle each datum to obtain better results? To answer these questions, the data problem is divided into three classes: (i) data sources, (ii) dimensions, and (iii) data management.

\subsubsection{Data Sources}
\label{DataS}

The type of datasets from the air transportation system are mainly related to airlines, airports or ensemble. Since airlines and airports commonly do not share their databases with the entire community, they are often used by collaborators of those institutions. Ensemble datasets may include both carriers, airports, and additional information provided by governmental agencies, regulatory authorities, and service providers. Table \ref{tsources} displays the type of datasets by regions. It presents the number of publications and the top three most cited papers in each category. Governmental agencies usually provide public access to their databases with different granularity. It is noticed that data from The United States Department of Transportation \cite{dot_united_2017}, primarily through The Federal Aviation Administration \cite{faa_federal_2017} and The Bureau of Transportation Statistics databases \cite{bts_bureau_2017} are widely used to obtain information about flights. The Eurocontrol \cite{eurocontrol_european_2017} database is provided by an intergovernmental organization in Europe. This dataset is also used intensively in flight delay studies \cite{reynolds-feighan_assessment_1999}. 

\begin{table}[!ht]
	\centering 
	\caption{Number of sources of real data about the air transportation system per region}
	\begin{tabular}{ C{2.5cm} C{2.5cm} C{2.5cm} C{2.5cm} }
		\hline\noalign{\smallskip}
		Region & Ensemble & Airline & Airport \\
		\hline\noalign{\smallskip}
		Asia & 2 \cite{mou_temporal_2017,takeichi_prediction_2017} & 1 \cite{rong_prediction_2015} & 1 \cite{xiangmin_departure_2017}\\
		Brazil & 2 \cite{sternberg_analysis_2016,anac_agencia_2017} & 0 & 0 \\
		Europe & 7 \cite{carr_evaluation_2005,campanelli_modeling_2014,lehouillier_measuring_2016} & 2 \cite{soomer_scheduling_2008,gurbuz_data_2011} & 7 \cite{reynolds-feighan_assessment_1999,bubalo_airport_2011,pejovic_factors_2009}\\
		US & 11 \cite{mueller_analysis_2002,tu_estimating_2008,zhang_hierarchical_2012} & 7 \cite{lan_planning_2006,ahmadbeygi_analysis_2008,ahmadbeygi_decreasing_2010} & 16 \cite{fleurquin_systemic_2013,balakrishna_accuracy_2010,fleurquin_data-driven_2013}\\
		\hline\noalign{\smallskip}
	\end{tabular}
	\label{tsources}
\end{table}

Other related datasets, such as weather, may be obtained from governmental databases or service providers. This includes, for example, The National Oceanic and Atmospheric Administration of the United States \cite{noaa_national_2017}. In fact, authors may use more than one source to develop their models. Datasets from United States Department of Transportation \cite{dot_united_2017}, National Oceanic and Atmospheric Administration \cite{noaa_national_2017}, and Weather Company \cite{twc_weather_2017} are commonly used to build delay prediction models. 

Additionally, some researchers \cite{zonglei_new_2008, zou_flight_2012} create synthetic datasets to evaluate their models instead of using real data. For example, Zou et al. \cite{zou_flight_2012} developed a market scenario, considering airport capacity, links, frequency, and characteristics of flights and passenger demand.

\subsubsection{Dimensions}
\label{Dimensions}

\begin{figure}[h]
	\centering\includegraphics[width=\linewidth]{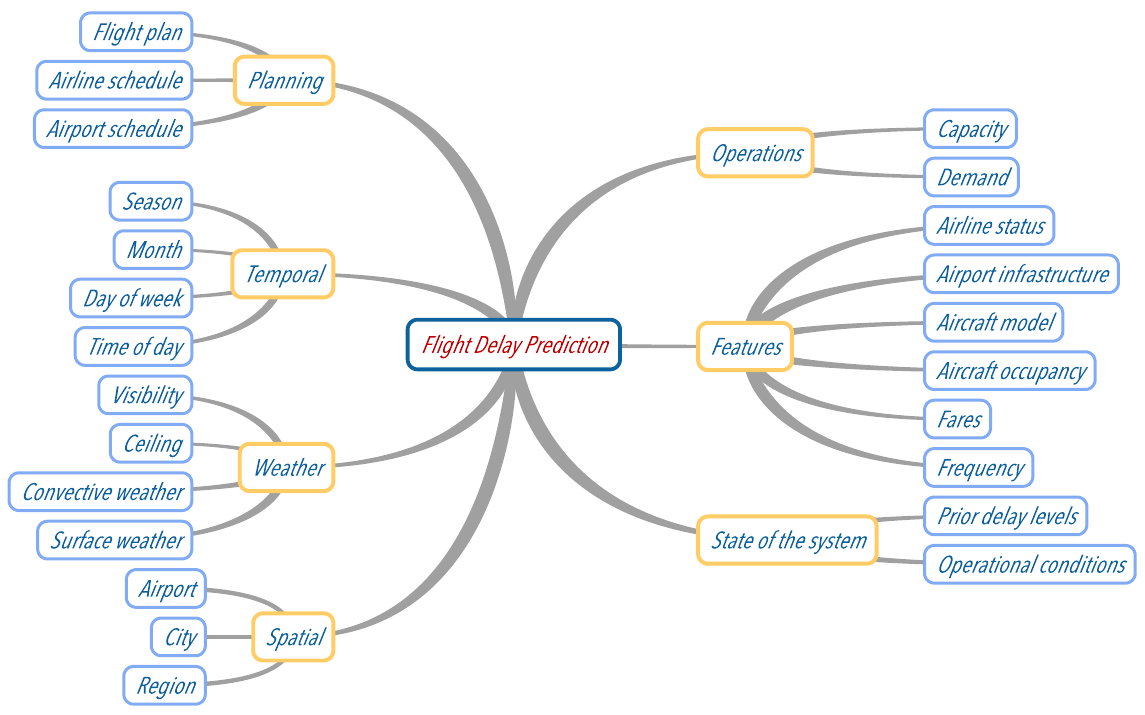}
	\caption{Data model of the flight delay prediction}
	\label{fig:Datamodel}
\end{figure}

Considering the main public datasets and the papers analyzed, we have organized them main commonly attributes used into seven classes depicted in the data model of Figure \ref{fig:Datamodel}. They abstract the main input attributes for delay prediction models. Beyond scheduled and actual times of departure and arrival, several characteristics may be considered depending on the focus of research. 

Spatial dimension is related to the positions taken by the aircraft, such as departure and arrival airports, their cities, regions, and countries \cite{hao_new_2014, rebollo_characterization_2014}. The temporal dimension is often used to capture seasonality or periodic patterns of data. These elements contain both date (season, month, and day of the week) and time (the day or time of the day) characteristics \cite{mueller_analysis_2002, abdel-aty_detecting_2007, tu_estimating_2008}. Weather dimension expresses external and environmental conditions in a particular moment \cite{evans_improving_2016}. It may represent specific features, such as ceiling and visibility \cite{reynolds-feighan_assessment_1999} that defines, for example, if take-off or landing is going to happen under visual or instrumental conditions. Additionally, \emph{en route} airspace weather situation (known as convective weather) and airport weather situation (known as surface weather) contain several momentaneous parameters \cite{hunter_advanced_2007}. 

Planning describes what airlines, airports, and air traffic controllers intend to do with critical resources involved in their operations. This dimension includes (i) airline schedules, (ii) airport schedules and (iii) flight plans. Arline schedules define all origin and destination points, their frequency and sequence, and aircrafts and crew allocations for each flight \cite{boswell_analysis_1997, beatty_preliminary_1998, wu_inherent_2005, ahmadbeygi_analysis_2008,duck_increasing_2012}. Airport schedules indicate the time each flight takes-off and lands, while flight plans indicate all \emph{en route} parameters, such as distance, route, speed, and high \cite{hansen_micro-level_2002}.

Features represent characteristics of airlines, airports or aircrafts. Airlines status may indicate if a carrier is a major or an affiliate one or if it is a traditional hub-and-spoke or a low-cost point-to-point. Aircrafts characteristics show their size, their number of seats and occupancy, which may be a constraint to some operations because they affect market decisions. Finally, airport infrastructure may represent the number of runways, gates and service providers in an airport facility \cite{pathomsiri_impact_2008,soomer_scheduling_2008,xiong_modelling_2013}. 

The state of the system indicates in which conditions airlines, airports or \emph{en route} airspace are operating at a specific moment. Some examples correspond to prior levels of delay or airports closures \cite{zonglei_new_2008}. The information about the state of the system is used to predict its behavior. Finally, operations are related to capacity and demand of airports and \emph{en route} airspace. When demand exceeds capacity, a congestion scenario is formed, which enables occurrence of delays \cite{morrison_effect_2008}. 

\subsubsection{Data Management}
\label{DataM}

Since the use of databases to store a massive amount of data have been increasing over the last years, data management techniques are becoming more and more crucial to provide a convenient and efficient query processing. Data management tasks contemplate design of database structure to enable data integration from different sources, elimination of inconsistencies, and data transformation. The development of a data warehouse supported by online analytical processing (OLAP) and data management techniques may be useful for this purpose. As mentioned in Section \ref{DataS}, multiple sources of data may be used. Thus, the usage of data warehouses combined with Extract, Transform and Load (ETL) procedures are commonly used to link the datasets of different sources \cite{yao_ria-based_2009}. 

There are many data management preprocessing procedures that can be applied to flight delay prediction datasets. They include data cleaning, feature selection, data transformation, and clustering. One of the main tasks of data cleaning is outlier removal. Extreme conditions may result in outliers that are not interesting if one is concerned about regular operations \cite{tu_estimating_2008}. Feature selection is the process of identifying attributes that are less correlated. Correlated and irrelevant attributes may provide model over-fitting or decrease prediction performance \cite{wong_survival_2012}. These preprocessing procedures are essential since the better the preprocessing is conducted on input data, the better the prediction models may be developed from it.

Data transformation is also an important activity to empower prediction models. Some examples of transformations include  normalization and discretization. Normalization reduces the range of possible values to a particular interval, such as -1 to 1 or 0 to 1. It gives equal strength for different variables and let machine learning methods identify which are the most relevant ones. Discretization consists of replacing numerical values by representative labels. It includes the transform of time periods into bins of a fixed time \cite{balakrishna_accuracy_2010, kim_deconstructing_2013}, binning of values to cope with limitations in computational packages \cite{boswell_analysis_1997, xu_estimation_2005} or to better train prediction models \cite{beatty_preliminary_1998}, especially when using machine learning models.

Clustering means grouping elements of the dataset in a way that similar observations stay together in the same group and dissimilar items stay in different groups. Many works compute clustering techniques, such as k-means or agglomerative hierarchical clustering, to support preliminary steps for further prediction models \cite{rebollo_characterization_2014}. 

\subsection{Method}
\label{Method}

\begin{figure}[h]
	\centering\includegraphics[width=\linewidth]{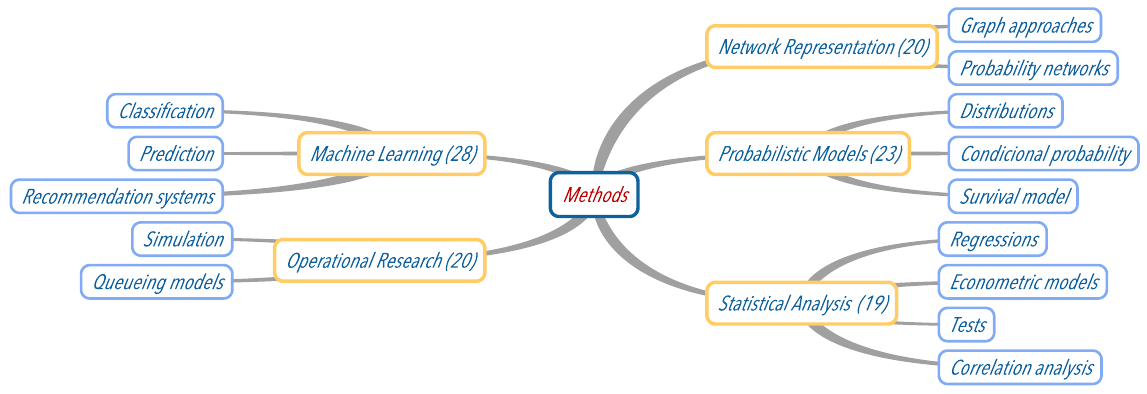}
	\caption{Categories of methods used to model the flight delay prediction}
	\label{fig:Categories}
\end{figure}

The flight delay prediction problem may be modeled in many ways, depending on the objectives of the research. Methods were divided into five groups, according to Figure \ref{fig:Categories}. The numbers next to each category represent the number of related papers.

\subsubsection{Statistical Analysis}
\label{StatisticalA}

Statistical analysis usually encompasses the use of regression models, correlation analysis, econometric models, parametric tests, non-parametric tests, and multivariate analysis (MVA). When it comes to regression models, both delay multiplier and recursive models can help airlines to understand delay propagation effects through the network and to estimate the costs of delays \cite{beatty_preliminary_1998, wang_flight_2003, markovic_statistical_2008,xu_multifactor_2008,zhang_airport_2010}. 

Many econometric models are also build to evaluate the efficiency flight systems, such as the analysis of the investments done by a governmental agency \cite{morrison_effect_2008} or to evaluate the equilibrium point considering the relationship between delays and passenger demand, fares, frequency and size of the aircrafts \cite{zou_flight_2012}. Xiong et al. \cite{xiong_modelling_2013} built an econometric model based on pre-existing delays, potential delay savings, distance, characteristics of the destination airport and airline, frequency, aircraft size, occupancy rate and fare to understand which reasons lead airlines to cancel their flights.  Qin et al. \cite{qin_statistical_2014} studied the periodicity of flight delay rate, whereas Mofokeng et al. \cite{mofokeng_factors_2017} studied the impact of aircraft turnaround time during maintenance check. Finally, Hao et al. \cite{hao_new_2014} built a model to quantify how delays originated at New York are propagated to other airports. 

Some works focus on statistical inference. Pathomsiri et al. \cite{pathomsiri_impact_2008} used a non-parametric function to evaluate the efficiency of airports of the United States regarding delays. Reynolds et al. \cite{reynolds-feighan_assessment_1999} computed the correlation between levels of delays and capacities of the European airports. They also suggested different approaches to deal with the congestion problem, describing their advantages and disadvantages. Finally, Abdel-Aty et al. \cite{abdel-aty_detecting_2007} calculated daily average of delays to detect correlations to understand the principal causes of delays at Orlando International Airport.

\subsubsection{Probabilistic Models}
\label{ProbabilisticM}

Probabilistic Models encompass analysis tools that estimate the probability of an event based on historical data. Tu et al. \cite{tu_estimating_2008} developed a probabilistic model based on expectation-maximization combined with genetic algorithms to predict the distribution of departure delay at Denver International Airport. 

Boswell et al. \cite{boswell_analysis_1997} expressed delay classes by a probabilistic mass function and used a transition matrix to verify delay propagation to subsequent flights. They made a cancellation analysis computing the conditional probability to cancel a flight given that its previous flight was delayed. Mueller et al. \cite{mueller_analysis_2002} modeled departure, \emph{en route} and arrival delays using density functions. The authors verified that Normal distribution fitted better to departure delays, while \emph{en route} and arrival delays were better described by Poisson distribution. Concerned about the total duration of a root delay, Wong et al. \cite{wong_survival_2012} studied delay propagation through a survival model. 

Evans et al. \cite{evans_modelling_2008} built a theoretical routing networks that integrated flight routing and scheduling model. Kotegawa et al. \cite{kotegawa_impact_2011} built a series of algorithms that forecast restructuring of the US commercial airline network to reduce both flight delay and total delay. Pfeil et al. \cite{pfeil_identification_2012} a probabilistic forecasts of whether or not a terminal area route will be blocked based on raw convective weather forecasts. Finally, Zhong et al. \cite{zhong_studies_2017} build a Monte Carlo simulations to estimate airports' runway capacity.

\subsubsection{Network Representation}
\label{NetworkR}

Network representation encompasses the study of flight systems according to a graph theory. Abdelghany et al. \cite{abdelghany_model_2004} built direct acyclic graphs to model the schedule of an airline (including flight times and resources availability) to detect disruptions and their impacts on the rest of the network. They used the classical shortest path algorithm to evaluate propagation effects. 

Ahmadbeygi et al. \cite{ahmadbeygi_analysis_2008} built propagation trees to compare two different airlines, one operating in a conventional hub-and-spoke scheme and the other in a low-cost point-to-point system. Xu et al. \cite{xu_estimation_2005} and Wu et al. \cite{wu_flight_2016} built a Bayesian network to model delay propagation.  Baspinar \cite{baspinar_data-driven_2016} built a network-epidemic process using historical flight-track data of Europe to create a novel delay propagation model. 

\subsubsection{Operational Research}
\label{OperationalR}

Operational Research includes advanced analytical methods (such as optimization, simulations, and queue theory) to help key-players make better decisions. Simulations may analyze airport capacity data, considering departure and arrival delays under different weather conditions \cite{schaefer_flight_2001, hunter_advanced_2007}. They may also evaluate the cost of, each delayed flight of an airline schedule \cite{soomer_scheduling_2008}. Moreover, simulations through queuing models were applied by Wieland \cite{wieland_limits_1997} to predict root delay, by Kim and Hansen \cite{kim_deconstructing_2013} to study the effects of capacity and demand on delay levels at the airports of New York area, and by Pyrgiotis et al. \cite{pyrgiotis_modelling_2013} to study delay propagation between some airports.

Other simulations were done to analyze delay propagation concerning schedule stability \cite{duck_increasing_2012} and reliability \cite{wu_inherent_2005}. Through simulations, different scenarios were commonly explored, such as reliability or flexibility of airports under external conditions. Hansen et al. \cite{hansen_micro-level_2002} considered the congestion problem and designed a simple deterministic queuing model to analyze propagation effects for subsequent flights of an airline and at Los Angeles International Airport.

\subsubsection{Machine Learning}
\label{MachineL}
Machine learning is the research that explores the development of algorithms that can learn from data and provide predictions based on it. Works that study flight systems are increasing the usage of machine learning methods. The methods commonly used include  k-Nearest Neighbor, neural networks, SVM, fuzzy logic, and random forests. They were mainly used for classification and prediction. 

Rebollo et al. \cite{rebollo_characterization_2014} applied random forests to predict root delay. They compared their approach with regression models to predict root delay in airports of the United States 
considering time horizons of 2, 4, 6 and 24 hours. Their test errors grew as the forecast horizon increased.

Khanmohammadi et al. \cite{khanmohammadi_systems_2014} created an adaptive network based on fuzzy inference system to predict root delay. The predictions were used as an input for a fuzzy decision-making method to sequence arrivals at JFK International Airport in New York.

Balakrishna et al. \cite{balakrishna_estimating_2008,balakrishna_accuracy_2010} used a reinforcement learning algorithm to predict taxi-out delays. The problem was modeled through a Markov decision process and solved by a machine learning algorithm. When running their model 15 minutes before the scheduled time of departure, authors achieved good performances at JFK International Airport in New York and Tampa Bay International Airport.

Lu et al. \cite{zonglei_new_2008} built a recommendation system to forecast delays at some airports due to propagation effects. The prediction was based on the k-Nearest Neighbor algorithm and used historical data to recognize similar situations in the past. The authors noticed fast response time and easy, logical comprehension as the main advantages of their method.

\section{Results and discussion}
\label{Results}

Since flight delays cause economic consequences to passengers and airlines, recognizing them through prediction may improve marketing decisions. Due to that, several forecast models have been built over the last twenty years. These models have sought to understand how delays propagate through the network of flights or airports, to predict root delay in the system or to comprehend the cancellation process. Beyond these three points of view for treating the flight delay prediction problem, models could also differ by their scope of application, data issues and methods.

\begin{figure}[h!t]
	\begin{minipage}[t]{0.5\textwidth}
		\null \centering
		\centering\includegraphics[width=1\linewidth]{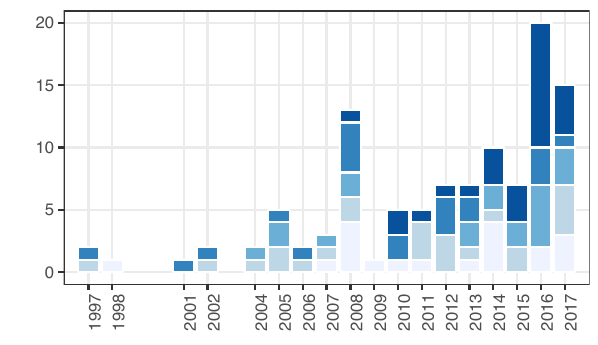} \\
		\centering a
	\end{minipage}%
	\hfill
	\begin{minipage}[t]{0.5\textwidth}
		\null \centering
		\centering\includegraphics[width=.8\linewidth]{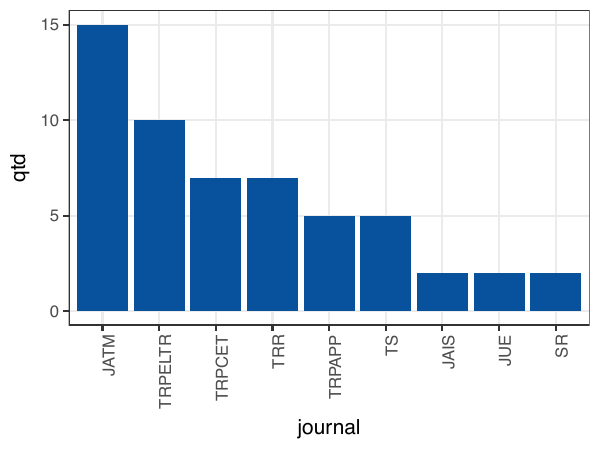} \\
		\centering b
	\end{minipage}
	\caption{(a) Publication in years according to main methods: $\STAT$~Statistical Analysis, $\PS$~Probabilistic Models, $\NET$~Network Representation, $\OR$~Operational Research, $\ML$~Machine Learning; \newline (b) Journals of with major published papers in the subject}
	\label{fig:publications_years}
\end{figure}

\begin{figure}[h!t]
	\centering 
	\small{
		\begin{tabular}{ C{1.5cm} C{5.5cm} C{5cm} }
			\hline\noalign{\smallskip}
			Years & root delay or cancellation & delay propagation\\
			\hline\noalign{\smallskip}
			1997-2000 & $\OR$~Wieland \cite{wieland_limits_1997} $\PS$~Boswell \cite{boswell_analysis_1997} & $\STAT$~Beatty \cite{beatty_preliminary_1998} \\
			2001-2004 & $\OR$~Hansen \cite{hansen_micro-level_2002} $\PS$~Mueller \cite{mueller_analysis_2002} $\PS$~Evans \cite{evans_quantifying_2004} & $\OR$~Schaefer \cite{schaefer_flight_2001} $\NET$~Abdelghany \cite{abdelghany_model_2004} \\
			2005 & $\NET$~Hsiao \cite{hsiao_air_2005} $\PS$~Hansen \cite{hansen_operational_2005} & $\OR$~Wu \cite{wu_inherent_2005} $\NET$~Xu \cite{xu_estimation_2005} \\
			2006 & $\PS$~Sim \cite{sim_potential_2006} & $\OR$~Lan \cite{lan_planning_2006} \\
			2007 & $\NET$~Wan \cite{wan_scalable_2007} $\PS$~Biesiada \cite{biesiada_gamma-ray_2007} $\STAT$~Abdel-Aty \cite{abdel-aty_detecting_2007} & \\
			2008 & $\ML$~Balakrishna \cite{balakrishna_airport_2008} $\OR$~Soomer \cite{soomer_scheduling_2008} $\NET$~McCrea \cite{mccrea_probabilistic_2008} $\PS$~Tu \cite{tu_estimating_2008} $\STAT$~Pathomsiri \cite{pathomsiri_impact_2008} & $\OR$~Lapp \cite{lapp_recursion-based_2008} $\NET$~AhmadBeygi \cite{ahmadbeygi_analysis_2008} \\
			2009 & $\STAT$~Pejovic \cite{pejovic_factors_2009} & \\
			2010 & $\ML$~Balakrishna \cite{balakrishna_accuracy_2010} $\OR$~Ganesan \cite{ganesan_improving_2010} $\STAT$~Klein \cite{klein_airport_2010} & $\OR$~Ahmadbeygi \cite{ahmadbeygi_decreasing_2010} \\
			2011 & $\ML$~Gürbüz \cite{gurbuz_data_2011} $\PS$~Evans \cite{evans_impact_2011} & $\STAT$~Nayak \cite{nayak_estimation_2011} \\
			2012 & $\ML$~Wang \cite{wang_prediction_2012} $\OR$~Zou \cite{zou_impact_2012} $\PS$~Azadian \cite{azadian_dynamic_2012} & $\OR$~Dück \cite{duck_increasing_2012} $\PS$~Wong \cite{wong_survival_2012} \\
			2013 & $\ML$~Kulkarni \cite{kulkarni_data_2013} $\OR$~Kim \cite{kim_deconstructing_2013} $\PS$~Evans \cite{evans_rebound_2013} $\STAT$~Xiong \cite{xiong_modelling_2013} & $\OR$~Pyrgiotis \cite{pyrgiotis_modelling_2013} $\NET$~Fleurquin \cite{fleurquin_systemic_2013} \\
			2014 & $\ML$~Rebollo \cite{rebollo_characterization_2014} $\PS$~Lin \cite{lin_border_2014} $\STAT$~Baumgarten \cite{baumgarten_impact_2014} & $\NET$~Campanelli \cite{campanelli_modeling_2014} $\STAT$~Hao \cite{hao_new_2014} \\
			2015 & $\ML$~Bloem \cite{bloem_ground_2015} $\NET$~Cai \cite{cai_novel_2015} $\PS$~Jacquillat \cite{jacquillat_endogenous_2015} & $\NET$~Ciruelos \cite{ciruelos_modelling_2015} $\PS$~Cheng \cite{cheng_hybrid_2015} \\
			2016 & $\ML$~Choi \cite{choi_prediction_2016} $\OR$~Castaing \cite{castaing_reducing_2016} $\NET$~Bertsimas \cite{bertsimas_unified_2016} $\STAT$~Simaiakis \cite{simaiakis_queuing_2016} & $\ML$~Khanmohammadi \cite{khanmohammadi_new_2016} $\NET$~Cong \cite{cong_empirical_2016} \\
			2017 & $\ML$~Takeichi \cite{takeichi_prediction_2017} $\ML$~Ding\cite{ding_predicting_2017} $\ML$~Baluch\cite{baluch_complex_2017} $\OR$~Jayam \cite{jayam_understanding_2017} $\PS$~Jacquillat \cite{jacquillat_dynamic_2017} $\STAT$~Pérez–Rodríguez \cite{perezrodriguez_modelling_2017} & $\NET$~Belkoura \cite{belkoura_beyond_2017} $\PS$~Ben Ahmed \cite{ben_ahmed_hybrid_2017} \\
			\hline\noalign{\smallskip}
		\end{tabular}
	}
	\caption{Time line of flight delay prediction publications:
		$\STAT$~Statistical Analysis, $\PS$~Probabilistic Models, $\NET$~Network Representation, $\OR$~Operational Research, $\ML$~Machine Learning}
	\label{fig:timeline}
\end{figure}

\begin{figure}[h!t]
	\centering\includegraphics[width=0.85\linewidth]{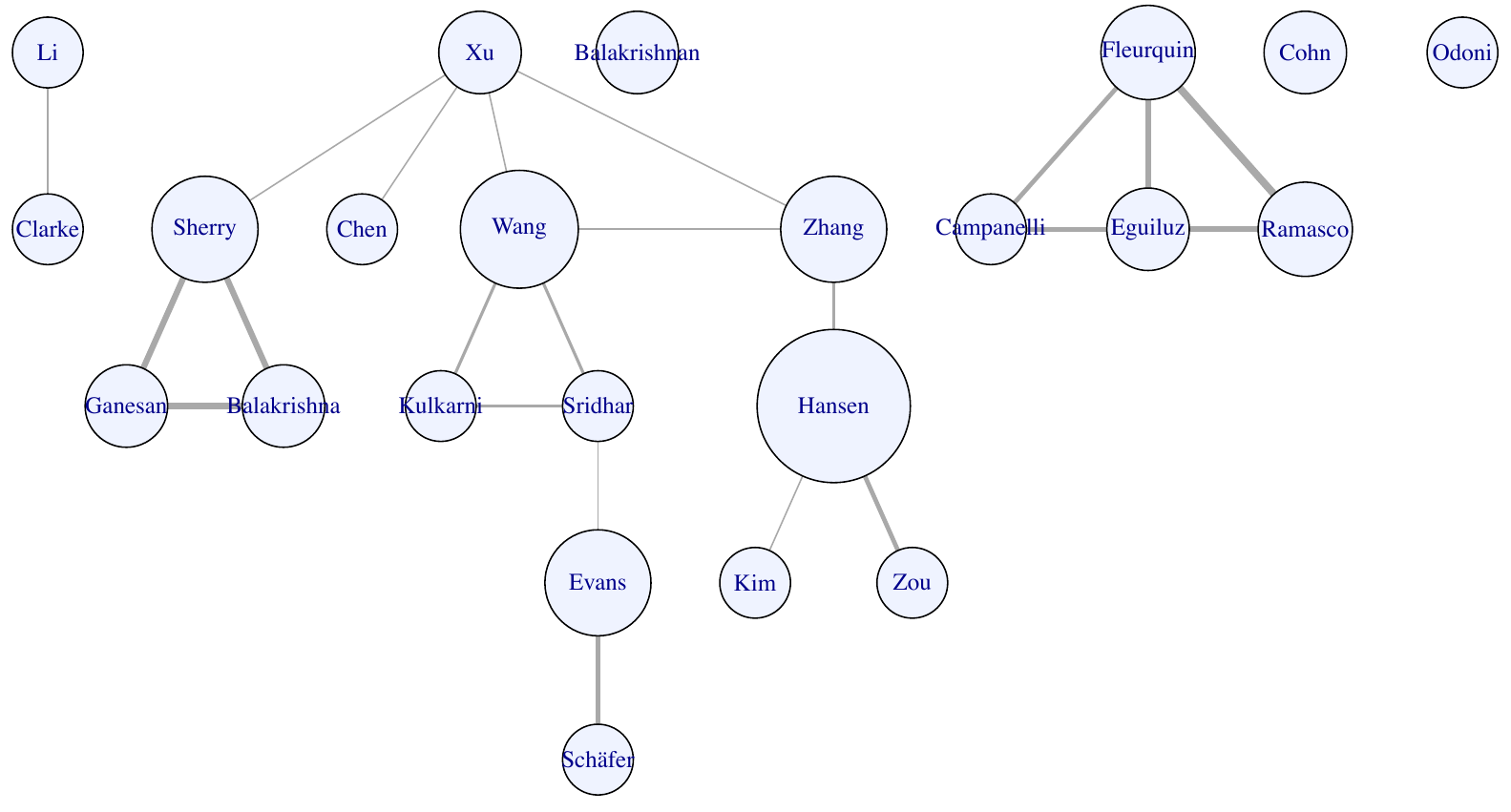}
	\caption{Collaboration network of main authors in subject}
	\label{fig:publications_network}
\end{figure}

The number of papers has increased in the late 2000s since 87.5\% of the works had been published between 2007 and 2017. Regarding only the documents considered in this analysis, Figure \ref{fig:publications_years}.a displays the number of publications grouped by methods. It can be observed a significant growth in machine learning \cite{ariyawansa_review_2016} and data mining \cite{belcastro_using_2016,kulkarni_analysis_2014} in the last decade. Also, Figure \ref{fig:timeline} depicts the complete timeline of papers, showing most cited authors per period and categories of methods. Pondering the way for tackling the delay problem, it was seen a balance between the number of papers that consider delay propagation and root delay, while few works deemed sole the cancellation analysis. Also, Figure \ref{fig:publications_years}.b indicates the foremost journals in which flight delay material was published.

From Figures \ref{fig:timeline} and \ref{fig:publications_network}, it is possible to observe the leading authors in the field. Figure \ref{fig:publications_network} displays the main collaboration graph from authors in our systematic review that had three or more publications. The radius of each vertex indicates the number of papers published by each author, whereas the strength of the edge indicates the degree of collaboration among the pair of authors. Some authors do not contain connected edges, meaning that none of their collaborators achieved three publications in our review.  

According to data perspective, we divided our analysis into three parts: data sources, dimensions and data management. From our review analysis, the adoption of data sources depends mostly on the country or region where the study has been taken place. For example, in China, most works were based on airport data, while in the United States the primary source was The United States Department of Transportation \cite{dot_united_2017}.

Dimensions were not directly related to the type of problem, but to the scope of application. This characteristic is notable in this case. Attributes such as weather, capacity, and demand were characteristics of airport or \emph{en route} airspace scopes. On the other hand, airlines schedules indicated scopes that considered airlines elements. It was also observed several ensembles of different dimensions, showing that prediction models may be improved through the selection of different attributes.

Data management was not specific to any problem or scope of application, and its use is steadily growing. In fact, it is present in most of the machine learning models adopted, primarily through data transformation. Most of the probabilistic models also considered outlier removal and data transformations techniques. A small percentage of the statistical analysis, network representation, and operational research methods applied general data management techniques as well.

\begin{figure}[h!t]
	\centering\includegraphics[width=0.5\linewidth]{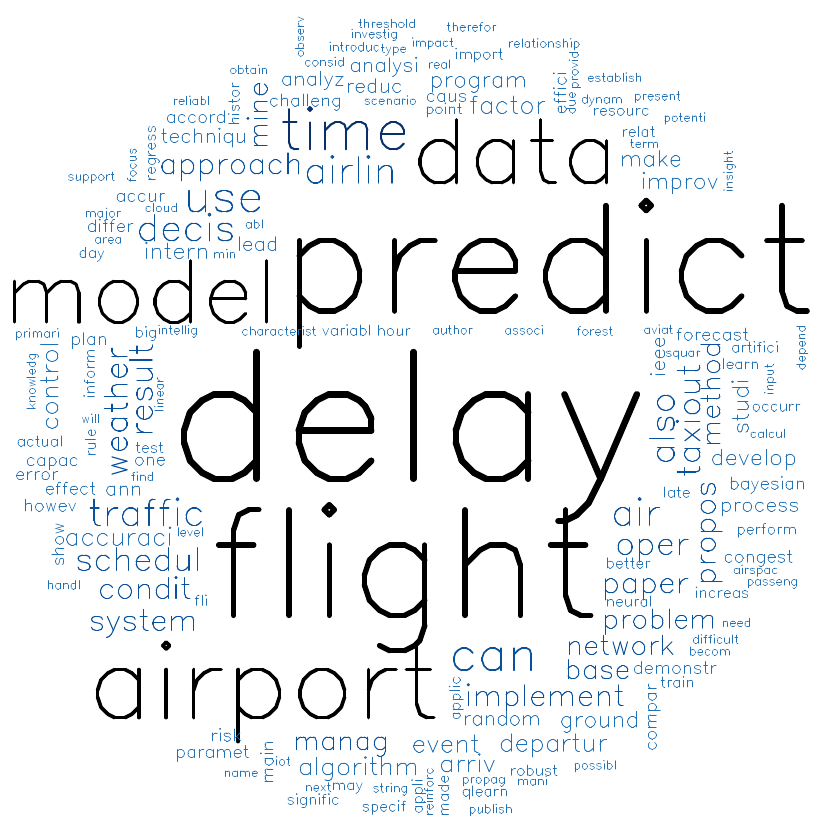}
	\caption{Trends in machine learning for flight delay prediction}
	\label{fig:publications_cloud}
\end{figure}

Regarding the methods used to develop the prediction models, statistical analysis, and operational research were the most applied in the past. These approaches were well spread between the three ways of treating the prediction problem. This same balance was also verified for probabilistic models. On the other hand, network representation was mostly employed for delay propagation. 

It is worth mentioning that machine learning approaches experienced a notable growth in the late 2000s, especially in root delay. In fact, both machine learning and data management are positively correlated. The more machine learning is used, the more data management is required. Especially, due to a trend in which extensive data is collected from sensors and IoT devices \cite{karakostas_event_2016,diana_validating_2014, peterson_economic_2013, xiong_modelling_2013, zhang_hierarchical_2012}. In fact, this can be confirmed in Figure \ref{fig:publications_cloud} that presents the cloud word from papers published between 2015 and 2017 related to flight delays and machine learning. Terms such as algorithm \cite{balakrishnan_control_2016}, big data \cite{cruciol_air_2015,cheng_risk_2016}, data model \cite{cox_ground_2016}, learn \cite{george_reinforcement_2016}, train-test \cite{ionescu_data_2016} are becoming more frequent. Such terminology is day-by-day becoming a trend for the next years.

\section{Conclusion}
\label{Conclusion}

Flight delays are an important subject in the literature due to their economic and environmental impacts. They may increase costs to customers and operational costs to airlines. Apart from outcomes directly related to passengers, delay prediction is crucial during the decision-making process for every player in the air transportation system.

In this context, researchers created flight delay models for delay prediction over the last years, and this work contributes with an analysis of these models from a Data Science perspective. We developed a taxonomy scheme and classified models in respect of detailed components.

Mainly, the taxonomy includes domain and Data Science branches. The former branch categorizes the problem (flight delay prediction) and the scope. The last branch groups methods and data handling. It was observed that the flight delay prediction is classified into two main categories, such as delay propagation and root delay and cancellation. Besides, the scope determines one of the three specific extents: airline, airport, en-route airspace or an ensemble of them. 

Additionally, considering Data Science branch, we aimed at the datum, by categorizing data sources, dimensions that can be used in the models, and data management techniques to preprocess data and improve prediction models efficiency. We also studied and divided the main methods into five categories: statistical analysis, probabilistic models, network representation, operations research, and machine learning. Those categories have been grouped as their use on specific forecast models for flight delays.

Besides the taxonomic scheme, we also presented a timeline with all articles to spot trends and relationships involving the main elements in the taxonomy. In the light of the domain-problem classification, this timeline showed a dominance of delay propagation and root delay over cancellation analysis. Researchers used to focus on statistical analysis and operational research approaches in the past. However, as the data volume grows, we noticed the use of machine learning and data management is increasing significantly. This clearly characterizes a Data Science trend.

Researchers from airlines, airports, and academia will require a combination of skills of both domain specialists and data scientists to enable knowledge discovery from flight Big Data. 

\section*{Acknowledgments}
The authors thank CNPq, CAPES (finance code 001), FAPERJ, and CEFET/RJ for partially funding this research.

\section*{Conflict of interest}
On behalf of all authors, the corresponding author states that there is no conflict of interest.

\bibliographystyle{abbrvnat}
\bibliography{references}

\end{document}